\documentclass[]{mn2e}

\usepackage{graphics}
\usepackage{rotating} 
\usepackage{amsfonts}

\def\apj{ApJ}      
\def\aap{A\&A}      

\def\apjs{ApJS}  
\def\aaps{A\&AS}  
\def\araa{ARA\&A} 

\def\mnras{MNRAS}        
         
\def\nat{Nature}
\def\pasp{PASP}

\def\ueber#1#2{{\setbox0=\hbox{$#1$}%
  \setbox1=\hbox to\wd0{\hss$ #2$\hss}%
  \offinterlineskip
  \vbox{\box1\box0}}{}}

\def\lesssim{\,\lower 1mm \hbox{\ueber{\sim}{<}}\,}
\def\grsim{\,\lower 1mm \hbox{\ueber{\sim}{>}}\,}

\title{The white dwarf luminosity  function --- II. The effect of the
       measurement errors and other biases}

\author[Torres et al.]{Santiago Torres$^{1,2}$,     
                       Enrique Garc\'{\i}a--Berro$^{1,2}$ and 
                       Jordi Isern$^{2,3}$\\ 
                       $^1$Departament de F\'\i sica Aplicada,
                       Escola Polit\`ecnica Superior de Castelldefels,   
                       Universitat Polit\`ecnica de Catalunya,\\  
                       Avda. del Canal  Ol\'\i mpic s/n,
                       08860 Castelldefels, Spain\\   
                       $^2$Institut d'Estudis Espacials de Catalunya,
                       c/Gran Capit\`a 2--4, Edif. Nexus 104, 
                       08034 Barcelona, Spain\\
                       $^3$Institut de Ci\`encies de l'Espai, CSIC,  
                       Campus UAB, Facultat de Ci\`encies, Torre C-5, 
                       08193 Bellaterra, Spain\\ }
\begin{document}

\maketitle

\begin{abstract} 
The  disc white  dwarf luminosity  function is  an important  tool for
studying the Solar neighbourhood, since it allows the determination of
several Galactic parameters,  the most important one being  the age of
the Galactic disc.  However, only the $1/\mathcal{V}_{\rm max}$ method
has  been employed so  far for  observationally determining  the white
dwarf  luminosity  function,  whereas  for other  kind  of  luminosity
functions several other methods  have been frequently used.  Moreover,
the procedures  to determine the  white dwarf luminosity  function are
not free of biases. These  biases have two different origins: they can
either be  of statistical nature  or a consequence of  the measurement
errors.  In a previous paper we  carried out an in--depth study of the
first category  of biases for several  luminosity function estimators.
In this  paper we  focus on the  biases introduced by  the measurement
errors and on the effects of  the degree of contamination of the input
sample  used to  build the  disc  white dwarf  luminosity function  by
different  kinematical populations.   To  assess the  extent of  these
biases  we  use a  Monte  Carlo  simulator  to generate  a  controlled
synthetic population and analyse the behaviour of the disc white dwarf
luminosity function for several assumptions about the magnitude of the
measurement errors and for several degrees of contamination, comparing
the  performances of  the most  robust luminosity  function estimators
under such conditions.
\end{abstract}


\section{Introduction}

White dwarf stars  are well studied objects from  both the theoretical
and  observational  point  of   view.  Hence,  the  disc  white  dwarf
luminosity  function   provides  us  with  an   invaluable  wealth  of
information  about  the  Solar neighbourhood.   Consequently,  several
important Galactic  parameters can  be derived from  the observational
white  dwarf luminosity  function.   Among these  parameters the  most
important ones are the age of the Galaxy (Winget et al. 1987; Garc\'\i
a--Berro et al.   1988; Hernanz et al. 1994; Richer  et al.  2000) and
the stellar formation rate (Noh  \& Scalo 1990; D\'\i az--Pinto et al.
1994;  Isern et  al.  1995;  Isern et  al.  2001).   Additionally, the
luminosity function of disc  white dwarfs provides an independent test
of the theory  of dense plasmas (Segretain et al.   1994; Isern et al.
1997).  Finally, the white dwarf luminosity function directly measures
the current  death rate of  low-- and intermediate--mass stars  in the
local disc, which also provides  us with an important tool to evaluate
stellar evolutionary sequences.

The advent of  large automated surveys --- like  the Sloan Digital Sky
Survey (York  et al.  2000; Stoughton  et al.  2002;  Abazajian et al.
2003,  2004; Eisenstein et  al.  2006),  the 2  Micron All  Sky Survey
(Skrutskie et al. 1997; Cutri et al. 2003), the SuperCosmos Sky Survey
(Hambly  et  al.  2001a;  Hambly,  et  al.   2001b; Hambly,  Irwin  \&
MacGillivray 2001), the 2dF QSO Redshift Survey (Vennes et al.  2002),
the SPY project (Pauli et  al.  2003), and others --- has dramatically
increased the number of  known white dwarfs.  Future astrometric space
missions --- of which {\sl Gaia} (Perryman et al. 2001) is the leading
example --- will undoubtedly increase  even more the size of the white
dwarf  population  with accurately  determined  parameters (Torres  et
al. 2005).  However,  this rapid increase in both  the quality and the
amount  of  observational  data   has  not  been  accompanied  by  the
corresponding  developements  in  the  way  in which  this  wealth  of
observational data  is analysed. Thus, there  is a need  to assess the
reliability of  the current  methods used to  estimate the  disc white
dwarf luminosity function  --- basically the $1/\mathcal{V}_{\rm max}$
method (Schmidt  1968) ---  and to test  other techniques  which allow
more  accurate determinations  of  the luminosity  function.  At  this
point it  is worth mentioning that new  luminosity function estimators
have  been  specifically   devised  to  solve  several  long--standing
problems for the case in  which galaxy luminosity functions have to be
obtained.  For instance, the C$^-$ method (Lynden--Bell 1971), the STY
method (Sandage  et al.  1979), the  Choloniewski method (Choloniewski
1986) and the Stepwise Maximum Likelihood method (Efstathiou, Ellis \&
Peterson et  al.  1988),  among other methods,  are currently  used to
derive  galaxy luminosity  functions.  Bivariate  luminosity functions
derived from the mixture  of two or more populations, non--homogeneity
biases  and  the  effects  of  anisotropies or  clustering,  are  some
examples of the kind of problems  that must be faced nowadays and that
these improved estimators are expected to correctly address.

Very few works have studied the reliability of the $1/\mathcal{V}_{\rm
max}$ method  when applied to the  case of the  white dwarf luminosity
function.  The  two preliminary studies  of Wood \& Oswalt  (1998) and
Garc\'\i a--Berro et al. (1999) demonstrated --- using two independent
Monte Carlo  simulators --- that the  $1/\mathcal{V}_{\rm max}$ method
for  proper--motion  selected samples  is  a  good density  estimator,
although it  shows important statistical  fluctuations when estimating
the slope  of the bright end  of the white  dwarf luminosity function.
In the  latter of these  works it  was also shown  that a bias  in the
derived ages of the solar neighbourhood is present, consequence of the
binning procedure.  Additionally it  has been recently shown (Geijo et
al. 2006)  that the size of  the observational error  bars assigned by
the  $1/\mathcal{V}_{\rm max}$ method  is severely  underestimated and
that  more robust luminosity  function estimators  can be  used. These
estimators provide a  good characterization of the shape  of the white
dwarf luminosity function even in the  case in which a small number of
objects is used.  Even more, in  this last study it was found that for
a small  sample size the  $1/\mathcal{V}_{\rm max}$ method  provides a
poor  characterization of  the less  populated bins,  while  for large
samples  the  performances  of  the  Choloniewski method  and  of  the
$1/\mathcal{V}_{\rm max}$  method are  very similar, providing  with a
reasonable accuracy both the shape  of the disc white dwarf luminosity
function and the precise location  of the cut--off.  Finally, the main
conclusion  obtained in  this  study was  that  in order  to obtain  a
reliable   observational   white   dwarf  luminosity   function   both
estimators, the $1/\mathcal{V}_{\rm  max}$ method and the Choloniewski
method  can  be   used,  while  other  parametric  maximum--likelihood
estimators are not recommended.

However, the approach adopted in  Geijo et al.  (2006) focused only on
the  statistical  techniques  used  to  obtain the  disc  white  dwarf
luminosity function,  whereas the effects of  the observational errors
and the contamination by different kinematical population were totally
disregarded.   The present  paper aims  precisely at  filling  in this
gap. Specifically, we study the  Lutz--Kelker bias --- see below for a
description of this  bias --- and the effects  of the contamination by
different kinematical  populations of the  input sample used  to build
the disc white  dwarf luminosity function.  To do this  we use a Monte
Carlo  simulator to  generate  a controlled  synthetic population  and
analyse the behaviour of the  disc white dwarf luminosity function for
several assumptions about the  magnitude of the measurement errors and
for several  degrees of  contamination, comparing the  performances of
the most robust luminosity function estimators under such conditions.

The paper  is organized as follows.   In section 2,  we briefly remind
the reader the basics of the  different estimators that can be used to
derive the  white dwarf  luminosity function, and  we argue  which are
best  fitted  for our  purpose.   In section  3  we  outline the  main
ingredients  of  the  Monte  Carlo  simulations  used  to  generate  a
synthetic population of white dwarfs  to which we apply the previously
described estimators.  A systematic study  of the effects on the white
dwarf  luminosity  function  of  the  measurement errors  and  of  the
contamination  by different  kinematical populations  is  performed in
section 4.  Finally, in section 5, we summarize our major findings and
we draw our conclusions.

\section{The luminosity function estimators}

The $1/\mathcal{V}_{\rm max}$ has been  the only method used up to now
for  observationally  determining  the  disc  white  dwarf  luminosity
function.  Since its introduction by  Schmidt (1968) in the studies of
the quasar population, it has  been extended to proper motion selected
samples  (Schmidt 1975)  and  generalized in  order  to introduce  the
dependence on the  direction of the sample (Felten  1976).  This turns
out  to be  useful when  studying  stellar samples  because the  scale
height of  the Galactic disc  introduces some biases.   Basically, the
$1/\mathcal{V}_{\rm max}$ method computes the maximum volume for which
a  star could  be a  member  of the  selected sample  given a  certain
proper--motion and magnitude limits.   The contribution of each object
to its  magnitude bin  is proportional to  the inverse of  its maximum
volume and the luminosity function  is built performing a weighted sum
over the  objects in  each magnitude bin.   Despite the fact  that the
$1/\mathcal{V}_{\rm  max}$   method  has  been   extensively  used  in
different  instances ---  it  has been  used  not only  to derive  the
luminosity function  of the  disc white dwarf  population but  also to
obtain luminosity functions of main sequence stars and quasars --- and
provides a reasonable estimate of the real luminosity function with an
easy computational implementation it also has important drawbacks. The
most   important  one   is   that   it  has   been   shown  that   the
$1/\mathcal{V}_{\rm  max}$ method should  only be  used when  both the
homogeneity  and  the  completeness  of  the sample  under  study  are
guaranteed.  This, obviously, is not an  easy task and for most of the
observational  samples  it  is   an  {\sl  ``a  priori''}  assumption.
Nevertheless, for the case of the white dwarf luminosity function this
is the  technique usually adopted for  observationally determining the
disc white dwarf luminosity function.

There  exist  other  alternatives  to  the  $1/\mathcal{V}_{\rm  max}$
method, mostly  based on a  maximum--likelihood analysis of  the data.
Among them,  the Choloniewski  method (Choloniewski 1986)  is probably
one of the most widely used.  The basic premise of this method is that
the local  distribution of  objects in some  pair of variables  of the
sample has a poissonian distribution.   Then, it is possible to define
a likelihood as  a function of the parameter  space.  The Choloniewski
method  divides the  parameter space  (magnitude and  parallax  in our
case) in  cells and  assumes poissonian statistics  for each  cell ---
see, for instance, Geijo et  al. (2006), and references therein, for a
complete  description  of   this  method.   Other  maximum--likelihood
estimators  can be  used, but  for the  case of  the disc  white dwarf
luminosity function the  Choloniewski method turns out to  be the most
appropriate one, as shown in Geijo  et al. (2006).  This is the reason
why in this  paper we will only compare the  performances of these two
methods.

\section{The Monte Carlo Simulations}

Since  we want  to study  the behaviour  of the  estimators previously
discussed  in \S  2  for the  realistic  case in  which two  different
kinematical  populations are  present, we  have built  synthetic white
dwarf  populations  in which  both  disc  and  halo white  dwarfs  are
generated using Monte Carlo  techniques.  We have thoroughly described
our Monte Carlo simulator in previous papers (Garc\'\i a--Berro et al.
1999; Torres et al.  2002; Garc\'\i  a--Berro et al.  2004) so here we
will only summarize the most important inputs.

We have used a  pseudo--random number generator algorithm (James 1990)
which  provides  a uniform  probability  density  within the  interval
$(0,1)$ and  ensures a  repetition period of  $\ga 10^{18}$,  which is
virtually infinite for  practical purposes.  When gaussian probability
functions are needed, we have used the Box--Muller algorithm (Press et
al.   1986).  Each  one of  the Monte  Carlo simulations  discussed in
section 4  consists of an  ensemble of 40 independent  realizations of
the synthetic  white dwarf  population, for which  the average  of any
observational quantity along with its corresponding standard deviation
were computed.  Here the standard deviation means the ensemble mean of
the sample dispersions for a typical sample.

We start  with the  disc model.  Firstly,  masses and birth  times are
drawn according to  a standard initial mass function  (Scalo 1998) and
an exponentially decreasing star  formation rate per unit surface area
(Bravo, Isern \& Canal 1993; Isern et al.  1995).  The spatial density
distribution is obtained from a  scale height law (Isern et al.  1995)
which varies  with time and  is related to the  velocity distributions
--- see below  --- and an exponentially decreasing  surface density in
the Galactocentric  distance.  The  velocities of the  simulated stars
are  drawn from  Gaussian distributions.   The  Gaussian distributions
take into account  both the differential rotation of  the disc and the
peculiar  velocity of  the Sun  (Dehnen  \& Binney  1997).  The  three
components  of the velocity  dispersion $(\sigma_{\rm  U}, \sigma_{\rm
V}, \sigma_{\rm W})$ and the lag velocity $V_0$ are not independent of
the scale  height but, instead, are  taken from the fit  of Mihalas \&
Binney (1981) to main sequence star counts.  It is important to notice
at this point that with this description we recover both the thick and
the thin disc  populations, and, moreover, we obtain  an excellent fit
to the disc white dwarf  luminosity function (Garc\'\i a--Berro et al.
1999).   The adopted  age of  the disc  is 11  Gyr. Finally,  the disc
simulations have  been normalized to  the local space density  of disc
white dwarfs within 250~pc, $n=0.5\times 10^{-3}$~pc$^{-3}$ for $M_V <
12.75^{\rm mag}$ (Liebert, Bergeron \& Holberg 2005).

For  the halo population  we adopt  a typical  isothermal, spherically
symmetric halo with a density profile given by the expression:

\begin{equation}
\rho(r)=\rho_0\frac{a^2+R_{\sun}^2}{a^2+r^2}
\end{equation}

\noindent where $a\approx  5$ kpc is the core  radius, $\rho_0$ is the
local  halo  density  and  $R_{\sun}=$8.5 kpc  is  the  Galactocentric
distance of the Sun. The velocity distributions are Gaussian:

\begin{equation}
f(v_r,v_{\theta},v_{\phi})=\frac{1}{(2\pi)^{3/2}}
\frac{1}{\sigma_r\sigma_t^2}\exp\left[-\frac{1}{2}\left(
\frac{v_r^2}{\sigma_r^2}+\frac{v_{\theta}^2+v_{\phi}^2}{\sigma_t^2}
\right)\right]
\end{equation}

\noindent   The  radial  and   tangential  velocity   dispersions  are
determined from  Markovi\'c \& Sommer--Larsen (1997).   For the radial
velocity dispersion we have:

\begin{equation}
\sigma_r^2=\sigma_0^2+\sigma_+^2\left[\frac{1}{2}
  -\frac{1}{\pi}\arctan\left(\frac{r-r_0}{l}\right)\right]
\end{equation}

\noindent where $\sigma_0=80\, {\rm km\,s^{-1}}$, $\sigma_+=145\, {\rm
km\,s^{-1}}$,  $r_0=10.5$   kpc  and  $l=5.5$   kpc.   The  tangential
dispersion is given by:

\begin{equation}
\sigma_t^2=\frac{1}{2}V_{\rm
c}^2-\left(\frac{\gamma}{2}-1\right)\sigma_r^2+   
\frac{r}{2}\frac{{\rm d}\sigma_r^2}{{\rm d}r}
\end{equation}

\noindent where
\begin{equation}
r\frac{{\rm d}\sigma_r^2}{{\rm d}r}=
-\frac{1}{\pi}\frac{r}{l}\frac{\sigma_+^2}{1+[(r-r_0)/l]^2}
\end{equation}

\noindent  For  the  calculations  reported  here we  have  adopted  a
circular velocity $V_{\rm  c}= 220$~km/s.  The halo was  assumed to be
formed in an  intense burst of star formation that  occured 14 Gyr ago
and lasted for 1 Gyr.

For both  sets of Monte Carlo simulations  the initial--to--final mass
relationship is  that of  Iben \& Laughlin  (1989). The  main sequence
lifetimes of the  progenitors of white dwarfs is  also taken from Iben
\& Laughlin (1989).  Finally, the  cooling sequences of Salaris et al.
(2000) have been  used.  These cooling sequences  incorporate the most
accurate   physical  inputs  for   the  stellar   interior  (including
neutrinos,  crystallization and  phase separation)  and  reproduce the
blue  turn at  low luminosities  (Hansen 1998).   Also,  these cooling
sequences  encompass the  range  of interest  of  white dwarf  masses,
therefore a complete  coverage of the effects of  the mass spectrum of
the white dwarf population was taken into account.

\section{Results}

\subsection{A sample with known parallaxes}

We  have  first  simulated  a  disc  white  dwarf  sample  with  known
parallaxes.  For  a realistic observational sample  this is equivalent
to saying that the maximum distance  for which a white dwarf can enter
into the sample  is $\sim 250$ pc. This is the  case, for instance, of
the catalog of spectroscopically  identified white dwarfs of McCook \&
Sion  (1999),  where  the  minimum  parallax  is  $\approx0.003\  {\rm
arcsec}$.  In  order to  build the final  sample from which  the white
dwarf luminosity  function is obtained,  we have chosen  the following
criteria:  $m_V\le 18.5^{\rm  mag}$ and  $\mu\ge 0.16^{\prime\prime}\;
{\rm yr}^{-1}$ as it was done in Oswalt et al.  (1996).  Additionally,
all white dwarfs  brighter than $M_V\le 13^{\rm mag}$  are included in
the sample,  regardless of their proper motions,  since the luminosity
function  of hot  white dwarfs  has been  obtained from  a  catalog of
spectroscopically identified white dwarfs  (Green 1980; Fleming et al.
1986) which is  assumed to be  complete (Liebert, Bergeron  \& Holberg
2005).  Moreover,  all white dwarfs with  tangential velocities larger
than 250 km~s$^{-1}$ were discarded (Liebert et al.  1989) since these
would be  probably classified as  halo members, according to  the most
widely  used  procedure.  Finally  we  have  added random  measurement
errors to  the proper motions,  apparent magnitudes and  parallaxes of
all  white dwarfs.  The  measurement errors  were drawn  from gaussian
distributions  with the  following deviations:  $\sigma_{\mu}=4\, {\rm
mas\;  yr^{-1}}$, $\sigma_{m_V}=0.02m_V$  and $\sigma_{\pi}=0.167\pi$,
which are realistic values (Luri et al. 1996; Harris et al. 2006).

\begin{figure}
\vspace{8.5cm}     
\includegraphics{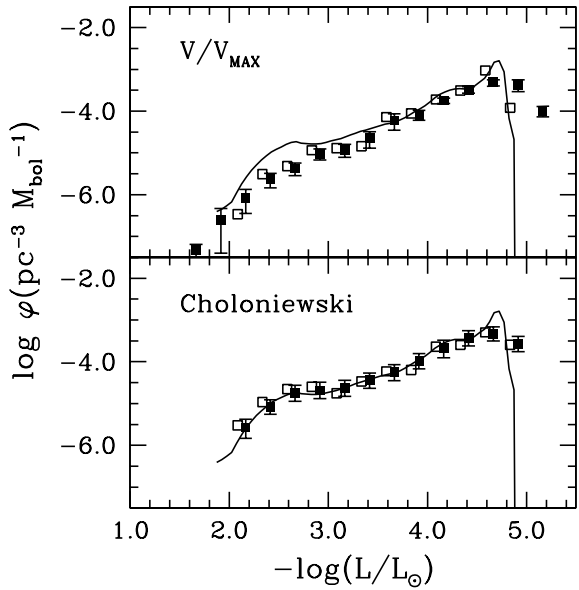}
\caption{White  dwarf luminosity  function for  a simulated  sample in
         which the white dwarfs have known parallaxes for the cases in
         which no measurement errors are considered (open symbols) and
         adding  proper motion, magnitude  and parallax  errors (solid
         symbols).  The open symbols  have been shifted by $\Delta\log
         (L/L_{\sun})=-0.08$ for the sake  of clarity.  The solid line
         shows the real luminosity function.}
\end{figure}

In Fig.  1 we show  the disc white dwarf luminosity functions obtained
using this method,  for both the case in  which the measurement errors
were disregarded (open symbols) and  the case in which the measurement
errors  were fully  taken  into account  (filled  symbols).  The  open
symbols have  been shifted by $\Delta\log  (L/L_{\sun})=-0.08$ for the
sake of  clarity. In  the upper panel  the results obtained  using the
$1/\mathcal{V}_{\rm max}$ method are shown whereas in the bottom panel
we display  the results obtained using the  Choloniewski estimator. We
recall that,  by construction, our  samples are complete,  although we
only  select  about 300  white  dwarfs  using  the selection  criteria
discussed  before.   However, our  simulations  do  provide the  whole
population of white dwarfs, which is much larger. Hence, we can obtain
the {\sl real} luminosity function  by simply counting white dwarfs in
the computational volume.  This is done for all realizations and there
after we obtain the average. The result is depicted as a solid line in
Fig.   1.   The  true   luminosity  function  steadily  increases  for
luminosities  larger  than  $\log(L/L_{\sun})\simeq-4.7$ and  then  it
decreases.   The  sharp  drop  at $\log(L/L_{\sun})\simeq-4.9$  is  an
artifact  of the  numerical  procedure because  no  white dwarfs  more
massive than $M_{\rm WD}\simeq 1.1\,M_{\sun}$ have been simulated. The
reason for this  is that no reliable cooling  sequences were available
until very recently (Althaus et al. 2007).

\begin{figure}
\vspace{8.5cm}     
\includegraphics{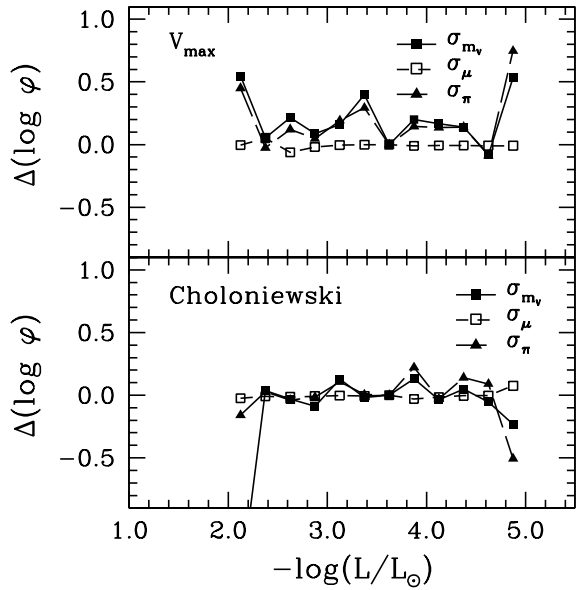}
\caption{Differences of the resulting white dwarf luminosity function,
         $\Delta \log \phi$,  when apparent magnitude (solid squares),
         proper  motion  (open  squares)  and parallax  errors  (solid
         triangles)  are  assumed, with  respect  to  the white  dwarf
         luminosity  function  with no  errors  added.   See text  for
         details.}
\end{figure}

We focus  first on  the overall shape  of the luminosity  function and
later we study  the position of the cut--off.  Both estimators recover
with a relatively good degree  of accuracy the slope of the increasing
branch  of the  disc  white dwarf  luminosity  function. However,  the
Choloniewski method performs  much better than the $1/\mathcal{V}_{\rm
max}$  estimator for  luminosities larger  than $\log(L/L_{\sun})=-3$.
Specifically,  the $1/\mathcal{V}_{\rm max}$  method shows  a markedly
tendency  to  underestimate  the  density  of  white  dwarfs  for  the
brightest luminosity  bins.  This  was already found  in Geijo  et al.
(2006),  and it  is a  bias  which can  be attributed  to the  limited
distance  of the  sample.   To be  precise,  for a  magnitude--limited
sample, bright white dwarfs can enter into the sample even if they are
located  at   larger  distances   than  faint  stars.    Although  the
$1/\mathcal{V}_{\rm  max}$  method has  been  devised  to remove  this
effect,  the   bias  persists  for   bins  with  a  small   number  of
objects. Note that for both  luminosity estimators, the effects of the
measurement errors are small and, in most of the cases, the luminosity
function in which the measurement errors were disregarded falls inside
the  $1\sigma$ error  bars of  the  case in  which we  have added  the
measurement errors.

We now pay attention to the position of the cut--off of the disc white
dwarf luminosity  function.  Both  estimators recover fairly  well the
position of  the cut--off of  the white dwarf luminosity  function for
the case in  which the measurement errors are  not taken into account.
However, the situation turns out  to be different when the measurement
errors are taken into account. As  seen in the bottom panel of Fig.  1
the Choloniewski  method also recovers  very well the position  of the
cut--off in this case  while the $1/\mathcal{V}_{\rm max}$ method does
not, as  clearly shown in the  upper panel of Fig.   1. In particular,
for  this case  the bins  around the  maximum appear  to  have smaller
number  densities  and the  cut--off  of  the  white dwarf  luminosity
function   is   shifted    to   considerably   smaller   luminosities.
Specifically,  the position  of  the cut--off  is  shifted by  $\Delta
\log(L/L_{\sun})\simeq-0.3$.   This bias in  the determination  of the
cut--off has dramatic consequences, since the determination of the age
of the  Solar neighbourhood relies  on a precise determination  of the
position  of  the  cut--off,  and  depending on  the  adopted  cooling
sequences  and main  sequence lifetimes  this systematic  effect could
amount to about 2 Gyr (Garc\'\i a--Berro et al. 1999).

We now ask  ourselves which is the respective  contribution of each of
the measurement  errors to  this behaviour. To  state this  in another
way:  which  measurement errors  are  responsible  for  this shift  to
smaller  luminosities of  the position  of the  cut--off of  the white
dwarf luminosity function?  To  answer this question we have performed
a series of calculations where  the measurement errors have been added
separately. The results are shown in Fig.  2.  For the sake of clarity
we   have  represented  the   logarithmic  difference,   $\Delta  \log
\varphi=\log \varphi^\prime-\log  \varphi$ of the  luminosity function
in  which  the  measurement  errors  were taken  into  account,  $\log
\varphi^\prime$, with  respect to the white  dwarf luminosity function
in which no measurement errors were considered, $\log\varphi$. A first
inspection of Fig.   2 reveals that the effects  of introducing errors
in  proper motion  are  totally  negligible whereas,  as  it could  be
expected, the  effects of the  parallax and apparent  magnitude errors
are dominant and tightly  correlated.  Additionally, these effects are
more important  for both the  brightest and faintest  luminosity bins,
whereas the intermediate luminosity  bins are not so largely affected.
While  both estimators  are affected  by the  measurement  errors, the
$1/\mathcal{V}_{\rm max}$ method shows  a stronger dependence on these
for  the faintest luminosity  bins when  compared to  the Choloniewski
estimator.  This behaviour corresponds to the expectations, given that
the  Choloniewski estimator  is a  maximum--likelihood  estimator and,
consequently, the effects  of a single object are  thus smaller.  Also
remarkable is the fact that  the effects of the measurement errors are
of opposite  sign.  Generally speaking,  it has been  shown previously
that when no  measurement errors are taken into  account, the tendency
of the $1/\mathcal{V}_{\rm max}$ method is to underestimate the number
density of white dwarfs in the less populated luminosity bins, whereas
for the Choloniewski estimator  the tendency is the opposite. However,
when the  measurement errors are  incorporated the $1/\mathcal{V}_{\rm
max}$  method  shows  a  less  marked tendency  to  underestimate  the
luminosity  function in these  bins and  the Choloniewski  method also
produces a result which is closer to the real one.

\subsection{The Lutz--Kelker bias}

The  error  in   the  determination  of  the  parallax   for  a  small
parallax--limited sample  implies another kind of  bias. First studied
by Lutz  \& Kelker (1973),  these authors showed  that if we  assume a
monotonically decreasing  distribution of true  parallaxes, the errors
would  therefore scatter  more stars  into the  sample  (with positive
errors) than  out of it (with  negative errors). Even  more, this bias
was demonstrated  to be  independent of the  lower parallax  limit and
also  of  the parallax  distribution.   The  main  consequence of  the
Lutz--Kelker  bias  for  a  monotonically decreasing  distribution  of
parallaxes is  that the observed parallax results,  on average, larger
than its true value.   Consequently, this overestimate of the parallax
translates into an  underestimate of the distance and,  hence, into an
underestimate  of  the  luminosity   of  a  given  star.   A  complete
discussion  of the  Lutz--Kelker bias  can be  found in  Smith (2003).
However, for our analysis we have closely followed the nomenclature of
Binney  \&  Merrifield (1998).   Our  main  goal  is to  quantify  the
seriousness of  the Lutz--Kelker bias  for the white  dwarf luminosity
function and  how do the two  estimators under study  perform for this
particular case.  We  have proceeded as follows.  We  need to evaluate
the probability  $P(\pi|\pi'){\rm d}\pi$ that  the true parallax  of a
given star, $\pi$, lies within  $(\pi, \pi+{\rm d}\pi)$ given that its
measured parallax is $\pi'$.  To do this we apply Bayes theorem:

\begin{equation}
P(\pi|\pi')=\frac{P(\pi'|\pi)P(\pi)}{P(\pi')}
\end{equation}

We do not  need to evaluate the prior  probability of the denominator,
since it only appears as  a normalization factor, but the numerator is
of fundamental importance. Following  Binney \& Merrifield (1998), and
after some elementary algebra, the next expression is obtained

\begin{figure}
\vspace{7.0cm} 
\includegraphics{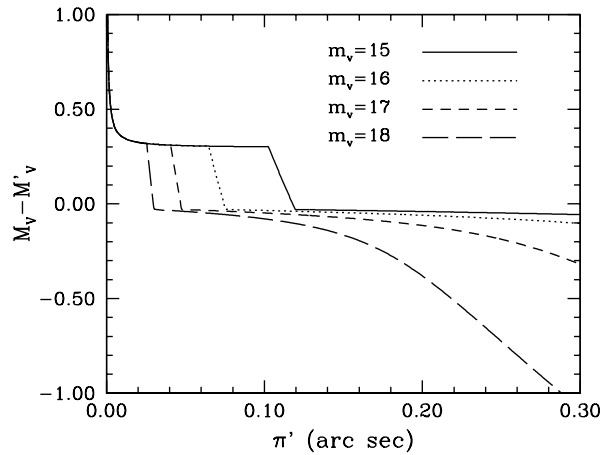}
\caption{Differences between the true absolute magnitude, $M$, and the
        magnitude inferred from the  raw parallax, $M'$, as a function
        of  the  measured   parallax  $\pi'$  for  different  apparent
        magnitudes $m$.}
\end{figure}

\begin{equation}
P(\pi|\pi')\propto P(\pi'|\pi)\Phi(M)\nu(s)\pi^{-4}
\label{bayes}
\end{equation}

\noindent where  $P(\pi'|\pi){\rm d}\pi'$ is the  probability that the
observational errors will cause the  parallax of a star, that has true
parallax $\pi$,  to be measured  as $\pi'$, $\Phi(M)$ is  the absolute
magnitude  luminosity function and  $\nu(s)$ is  the space  density of
stars with $s=\pi^{-1}$. In order to evaluate the Lutz--Kelker bias we
must   make   some   assumptions   with  respect   to   the   previous
functions. First of all, we assume that the probability of measuring a
parallax $\pi'$ for an object whose true parallax is $\pi$ is given by
a Gaussian  distribution with a standard  deviation, $\sigma_{\pi}$ of
the measurement errors.  For  the space density distribution and since
parallaxes can only be measured  for nearby stars, it is reasonable to
assume  that the distribution  $\nu$ is  independent of  $s$. Finally,
regarding the  luminosity function $\Phi(M)$ we can  take advantage of
our synthetic  population since  we know the  {\sl true} shape  of the
luminosity function.  We  must remark here that this  would not be the
case of a  real situation, where the luminosity  function is not known
{\sl  ``a  priori''}.   However,  this  problem can  be  solved  in  a
practical  situation by using  an iterative  procedure.  Additionally,
and for  the case of the  white dwarf luminosity  function, the bright
portion of the white dwarf  luminosity function turns out to be rather
insensitive  to  the  star   formation  rate,  which  also  eases  the
calculation  of  the  luminosity  function.  Under  these  assumptions
Eq. (\ref{bayes}) can be written in the following way:

\begin{equation}
P(\pi|\pi')\propto
\Phi(M)\pi^{-4}\exp\left[-\frac{(\pi'-\pi)^2}{2\sigma_{\pi}^2}\right]
\label{final}
\end{equation}

\noindent where $M=m+5\log(\pi/10)$.   For any given measured apparent
magnitude,  $m$,  and  parallax,  $\pi'$,  the  value  of  $\pi$  that
maximizes the previous  expression can be found.  It  is reasonable to
take this  value as the true  value of the parallax  and then estimate
the true absolute magnitude, $M=M'+5\log(\pi/\pi')$. In Fig. 4 we show
the  results  obtained  with   this  procedure  for  several  apparent
magnitudes,  ranging from  15 to  18, as  a function  of  the measured
parallax.   As   seen  in  this  figure,  for   small  parallaxes  the
Lutz--Kelker bias implies an  underestimate of the absolute magnitude,
that could be as large as  $1^{\rm mag}$.  On the contrary, the effect
for large  parallaxes is the  opposite.  The magnitude can  be largely
overestimated for faint white dwarfs.   It is worth noticing here that
this estimate  of the  true absolute magnitude  is independent  of the
luminosity  function estimator  and  only depends  on the  assumptions
adopted for  evaluating Eq. (\ref{final})  and, in particular,  on the
adopted luminosity function.

\begin{figure}
\vspace{10cm} 
\includegraphics{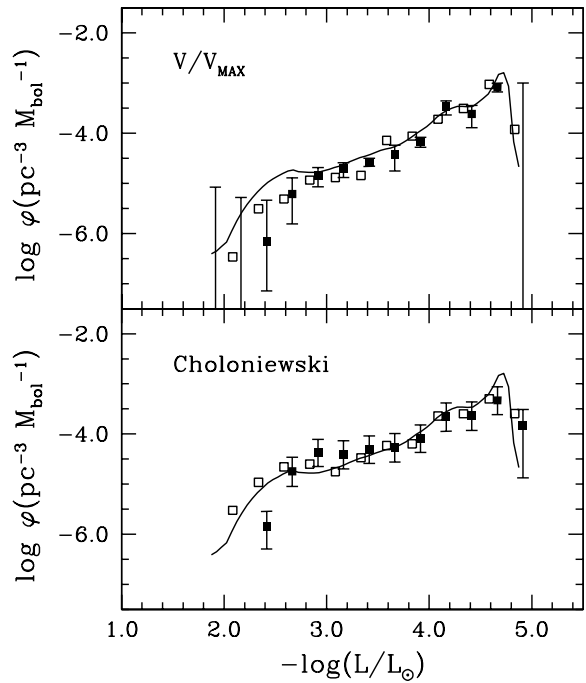}
\caption{White dwarf  luminosity function for a  simulated sample with
        known parallaxes assuming no measurement errors (open symbols)
        and for the  case in which measurement errors  were taken into
        account but after correcting  for the Lutz--Kelker bias (solid
        symbols).}
\end{figure}

Finally,  in Fig.   4 we  show  the resulting  white dwarf  luminosity
functions  derived using either  the $1/\mathcal{V}_{\rm  max}$ method
(top  panel)  or  the  Choloniewski  estimator  (bottom  panel)  after
correcting for the Lutz--Kelker  bias (solid symbols).  As before, our
real  luminosity function  is  represented  as a  solid  line and  the
luminosity function  in which no measurement  errors were incorporated
is  shown for the  sake of  comparison as  open symbols.   This figure
shows that  the effects of the  Lutz--Kelker bias can  be easily taken
into  account  and that  once  this is  done  we  recover the  correct
position of  the cut--off.  For the  rest of the bins  the effects are
not noticeable,  as one should expect.  Note as well that  the size of
the assigned error bars is  considerably larger when compared to those
obtained  previously for  both the  luminosity bins  of  the brightest
portion of the white dwarf luminosity function and for the bins at its
faint end.

\subsection{The SDSS simulation}

The very  recent publication  of the Sloan  Digital Sky  Survey (SDSS)
Data Release 3 has notably  increased the number of known white dwarfs
and  has  also  largely  extended  the search  volume  (Eisenstein  et
al. 2006).   Moreover, the SDSS sample, combined  with improved proper
motions from the USNO-B has  allowed to derive a preliminary (although
very  much   improved)  white  dwarf  luminosity   function  based  on
approximately 6000 stars  (Harris et al.  2006). Both  facts make this
sample an ideal testbed for the  the kind of techniques we are dealing
with.   However, we  recall  here that  the  SDSS does  not provide  a
parallax measurement  and, hence,  the distances are  determined using
the SDSS photometry.  In particular,  the distance is obtained using a
color--magnitude   relationship  for   an  otherwise   typical  $0.6\,
M_{\sun}$ white dwarf.  Taking this  into account, we have simulated a
disc white dwarf population centered around the North Galactic Cap, up
to a distance of 1800 pc  and according to the precise geometry of the
SDSS.   For  this simulation  we  have used  a  new  set of  selection
criteria, which  meet the characteristics  of the SDSS (Harris  et al.
2006).   These  selection   criteria  are  the  following:  $15.0^{\rm
mag}<m_V<19.5^{\rm  mag}$, $\mu>20\,  {\rm mas}\;  {\rm  yr}^{-1}$ and
$V_{\rm tan}>30\;  {\rm km\;s^{-1}}$. Harris  et al. (2006)  also used
the reduced  proper motion $H_g=g+5\log\mu+5$,  where $g$ is  the SDSS
magnitude,  to  discriminate between  main  sequence  stars and  white
dwarfs, since  the latter are typically 5--7  magnitudes less luminous
than subdwarfs  of the same  color, and this  is what we also  do.  As
previously done,  we have added the  corresponding measurement errors.
However, in  this case,  and given that  the parallax is  not directly
measured we have  only added proper motion and  magnitude errors.  The
distributions  of errors  are again  assumed to  be gaussian,  and the
corresponding  standard deviations  are $\sigma_{\mu}=4\,  {\rm mas}\;
{\rm  yr}^{-1}$  in  proper  motion, and  $\sigma_{m_V}=0.02\,m_V$  in
apparent magnitude.   The final  size of the  sample is  roughly $\sim
2000$  white dwarfs,  which is  very similar  to the  number  of white
dwarfs used by Harris et al. (2006).

\begin{figure}
\vspace{8.5cm}     
\includegraphics{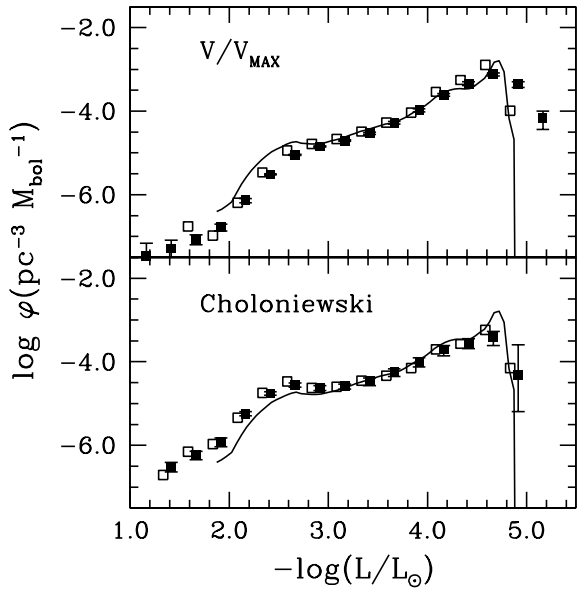}
\caption{White   dwarf  luminosity  functions   for  the   SDSS  model
         simulation assuming no  measurement errors (open symbols) and
         adding   proper   motion    and   magnitude   errors   (solid
         symbols). The solid line represents the {\sl real} luminosity
         function.}
\end{figure}

The  resulting white  dwarf luminosity  functions for  the  SDSS model
simulation assuming  no measurement  errors (open symbols)  and adding
proper motion and magnitude errors (solid symbols) are shown in Figure
5. We have  represented once again  the true luminosity function  as a
solid line. As seen, both estimators recover quite well both the shape
of the increasing  branch of the luminosity function  and the position
of the cut--off when no measurement errors are added.  For the case of
the $1/\mathcal{V}_{\rm max}$ method, the bias already mentioned for a
distance--limited sample, consisting in underestimating the density of
bright objects,  is somewhat reduced, whereas  the Choloniewski method
slightly  overestimates the  star density  for these  luminosity bins.
However, both  methods yield satisfactory results, and  we remark that
these  are minor effects.   However, when  the measurement  errors are
introduced, the  position of the cut--off  for the $1/\mathcal{V}_{\rm
max}$ estimator is again  shifted to smaller luminosities, whereas the
Choloniewski  method  correctly  retrieves  the  real  cut--off.   The
reasons  are   identical  to   those  previously  discussed   for  the
distance--limited sample --- the  Lutz--Kelker bias --- and, thus, the
Choloniewski method, which  is less sensitive to this  bias, turns out
to be more robust when analyzing this sample.

For  the  sake  of  completeness   we  have  also  evaluated  how  the
measurement errors  in magnitude  and proper motion  separately affect
the white  dwarf luminosity function.   The results, presented  as the
logarithmic difference  of the luminosity function, are  shown in Fig.
6.  The  behaviour  of  both  estimators  is quite  similar  to  those
presented if Fig.  2.  As seen, the error in apparent magnitude is the
dominant source  of errors for  the faintest luminosity bins  when the
$1/\mathcal{V}_{\rm  max}$ estimator  is used,  whereas the  errors in
proper  motion  play  a  very   limited  role  ---  as  it  should  be
expected. This is  also the case when the  Choloniewski method is used
but, for  the case of the  lowest luminosity bins, the  effects are in
any  case smaller than  when the  $1/\mathcal{V}_{\rm max}$  method is
employed.

\begin{figure}
\vspace{8.5cm}
\includegraphics{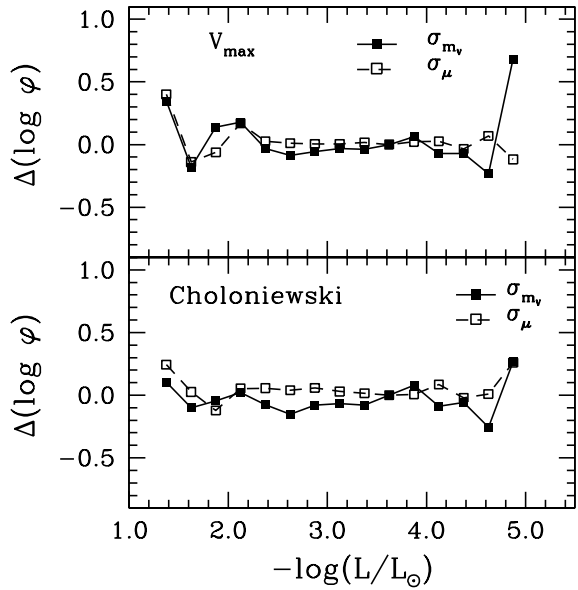}
\caption{Differences of the resulting white dwarf luminosity function,
         $\Delta \log  \phi$, when apparent  magnitude (solid squares)
         and proper motion (open squares) are assumed, with respect to
         the  white dwarf  luminosity function  with no  errors added.
         See text for details.}
\end{figure}

\subsection{Contamination by halo white dwarfs}

Up  to now  we have  been dealing  with model  white  dwarf luminosity
functions derived  from synthetic populations which  were derived from
pure kinematical populations and,  more particularly, from a sample of
disc white dwarfs. However, there  is some evidence that the faint end
of  the white  dwarf luminosity  function contains  multiple kinematic
populations, not only thin disc, but  also thick disc and halo as well
(Reid 2004).  Although  our model for the disc  white dwarf population
naturally incorporates the thick disc population, a careful evaluation
of  the effects  of the  contamination  by the  halo population  still
remains to be  done.  If, as expected, the sample  from which the disc
white dwarf  luminosity function is  built is contaminated by  a small
number of missclasified halo  white dwarfs, this may have consequences
on the location  of the observed cut--off. However  this still remains
to be  assessed.  In order  to clarify the  extent of such  a possible
bias and to  assess which is the response  of the different estimators
to this bias,  we have perfomed a series of  simulations where we have
introduced a certain degree of  halo contamination. Since the SDSS has
provided us with a large number  of new white dwarfs it is more likely
that a small degree of halo contamination should be present and, thus,
we assess  its effects using this last  simulation.  Additionally, and
in order  to see the real effects  of a small admixture  of halo white
dwarfs in  the resulting population, independently of  the role played
by  the measurement  errors,  we  use the  simulation  in which  these
measurement errors have been disregarded.

\begin{figure}
\vspace{7.0cm} 
\includegraphics{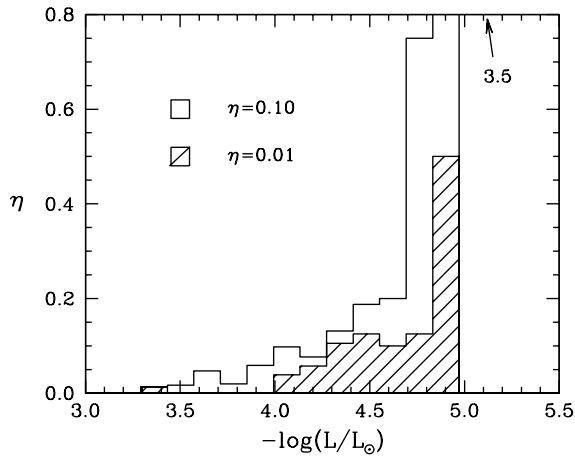}
\caption{Contamination of  the sample of  disc white dwarfs  with halo
         stars. The histograms show  the fraction of halo white dwarfs
         with respect to the total number of stars for each luminosity
         bin  as a function  of the  luminosity. The  shaded histogram
         corresponds to  a total contamination  of the sample  of 1\%,
         and  the  non--shaded   histogram  corresponds  to  a  global
         contamination of 10\%. See text for details. }
\end{figure}

We proceed as  follows. We simulate a population  of halo white dwarfs
within a total radius of 1800  pc. Then, we normalize the total number
of halo white dwarfs using the halo white dwarf luminosity function of
Torres et  al. (1998).  That  is, we impose  that the density  of halo
white dwarfs in the local neighbourhood (250 pc) is $n \sim 1.2 \times
10^{-5}$  pc$^{-3}$ for  $\log(L/L_{\sun})  \ga -3.5$  (Torres et  al.
1998).   We  then extract  all  halo white  dwarfs  which  are in  the
direction of the  region surveyed by the SDSS.  After this we randomly
include  the  selected  synthetic   halo  white  dwarfs  in  the  disc
sample. This  results in total  contamination of about  1\%.  Although
the total contamination turns out to be small, not all luminosity bins
are  equally  affected,  as  seen  in Fig.   7.   In  particular,  the
contamination of  the brightest luminosity  bins is moderate,  but for
the faintest luminosity bin the  contamination of the disc sample with
halo white dwarfs can be as large as 50\% --- as it should be expected
given that we have assumed that  the halo star formation history was a
burst of  very short duration which  occurred 14 Gyr  ago (see below).
Additionally,  and since  the local  density of  halo white  dwarfs is
still the  subject of  a strong  debate, we have  also adopted  a much
larger  total density  of  halo white  dwarfs  of $n  \sim 2.2  \times
10^{-4}$  pc$^{-3}$, as suggested  by Oppenheimer  et al.   (2001). We
stress that this  value of the density of halo  white dwarfs should be
considered  as  an extreme  upper  limit.   This  results in  a  total
contamination  of the  disc sample  with  halo white  dwarfs of  $\sim
10\%$.  However, as it was the case in which a modest contamination of
the  order 1\%  was discussed,  not  all luminosity  bins are  equally
affected.  In fact,  in this case the effects  are much more dramatic,
as clearly  seen in Fig. 7,  and we find that  the faintest luminosity
bins of the  white dwarf luminosity function are  totally dominated by
halo white  dwarfs.  Specifically, for the faintest  luminosity bin we
find  that halo white  dwarfs outnumber  disc members  by a  factor of
$\sim 3.5$, whereas at  luminosities of the order of $\log(L/L_{\sun})
\simeq  -4.0$,  the  degree  of  contamination  can  be  as  large  as
$\sim20$\%.
 
\begin{figure*}
\vspace{10cm} 
\includegraphics{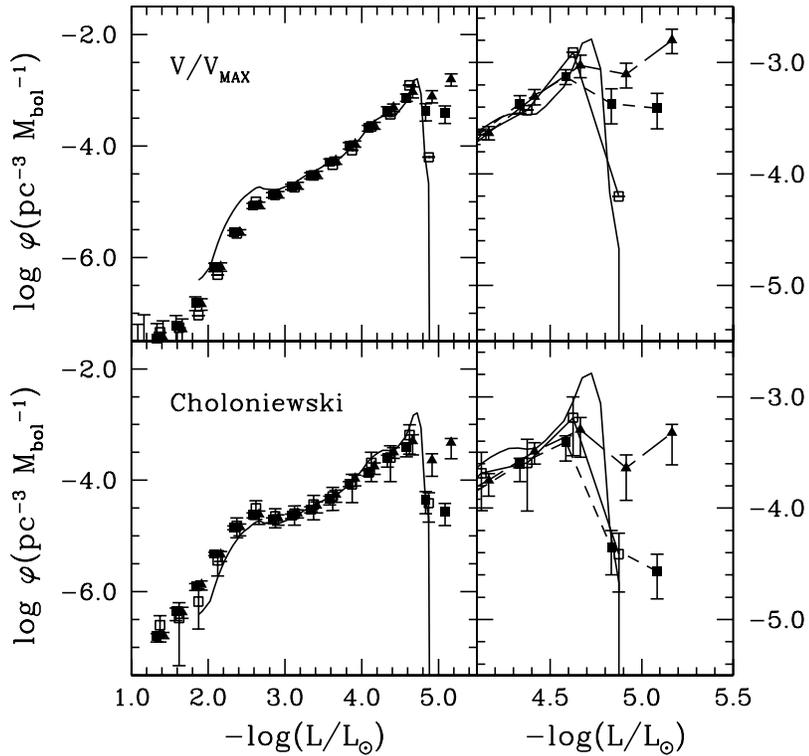}
\caption{The white  dwarf luminosity function for  the SDSS simulation
         when a  1\% of halo contamination is  assumed (solid squares)
         and assuming a 10\%  of halo contamination (solid triangles).
         The uncontaminated  white dwarf luminosity  function is shown
         as open  squares. The  top panels show  the results  when the
         $1/\mathcal{V}_{\rm  max}$  estimator  is used,  whereas  the
         bottom  panels  depict the  situation  when the  Choloniewski
         method is used. The right panels show an enlarged view of the
         region of low luminosities. See text for details. }
\end{figure*}

Now  we  ask  ourselves which  are  the  effects  of such  degrees  of
contamination in the resulting white dwarf luminosity function and how
the different estimators perform  in retrieving the correct luminosity
function. To  do this we  repeat the procedure  outlined in \S  4.3 to
select the  sample from which  the white dwarf luminosity  function is
built,  using  the   same  selection  criteria  previously  discussed:
$15.0^{\rm  mag}<m_V<19.5^{\rm  mag}$,   $\mu>20\,  {\rm  mas}\;  {\rm
yr}^{-1}$ and $V_{\rm tan}>30\; {\rm km\;s^{-1}}$. The overall results
are  shown in the  left panels  of Fig.   8.  In  this figure  we have
represented  the uncontaminated disc  white dwarf  luminosity function
(open  symbols), the  resulting disc  white dwarf  luminosity function
when a small  admixture of $1\%$ of halo white dwarfs  is added to the
previous white  dwarf population  (solid squares), and  the luminosity
function obtained when an admixture  of $10\%$ of halo white dwarfs is
added  to  the  uncontaminated  disc  white  dwarf  population  (solid
triangles).  For the sake of clarity, the right panels of Fig.  8 show
an expanded  view of  the region of  low luminosities.   As previously
done, the  top panels represent  the white dwarf  luminosity functions
obtained  using the $1/\mathcal{V}_{\rm  max}$ estimator,  whereas the
results obtained using the Choloniewski method are shown in the bottom
panels.  

As  it   should  be  expected   from  the  previous   discussion,  the
contamination by halo  white dwarfs only affects the  faintest bins of
the disc white dwarf luminosity function. This is so because since all
halo white  dwarfs are almost  coeval the halo white  dwarf luminosity
function  is strongly  peaked.  However,  for reasonable  ages  of the
stellar halo, the peak of  the halo white dwarf luminosity function is
located at luminosities much smaller  than that of the location of the
cut--off of  the disc white dwarf  luminosity function.  Consequently,
the  relative   contribution  of  halo  white   dwarfs  increases  for
decreasing  luminosities.   It   also  quite  apparent  the  different
behaviour of  both estimators. As can  be seen in the  right panels of
Fig. 8, even a small degree  of halo contamination --- of the order of
only $\sim 1\%$,  which corresponds to the halo  white density derived
by Torres  et al.  (1998) ---  strongly affects the shape  of the disc
white   dwarf  luminosity  function   for  the   case  in   which  the
$1/\mathcal{V}_{\rm  max}$  estimator is  used.   In  fact, the  sharp
drop--off in  the number density  of white dwarfs  which characterizes
the faint end of the white dwarf luminosity function is substituted by
a  shallow decrease  when the  degree of  contamination is  1\%  and a
moderate increase when  the degree of contamination is  10\% --- which
corresponds to the halo white  dwarf density derived by Oppenheimer et
al. (2001).  However,  the effects of the contamination  by halo white
dwarf are far  less apparent when the Choloniewski  estimator is used,
as clearly  shown in Fig.  8.   If the Choloniewski method  is used to
derive the white dwarf luminosity function the correct position of the
cut--off of the white dwarf luminosity function is still obtained when
the  degree  of  contamination  is  of  1\%,  although  an  additional
luminosity bin is obtained. On the contrary, when the contamination by
halo white  dwarfs is of the order  of 10\% the drop--off  of the disc
white dwarf  luminosity function is substituted by  a shallow decrease
and at even fainter luminosities the luminosity function increases, as
it was the  case when $1/\mathcal{V}_{\rm max}$ method  was used.  All
in all,  we find  that the  Choloniewski method turns  out to  be more
robust against a possible contamination by halo white dwarfs.

\begin{figure}
\vspace{10cm} 
\includegraphics{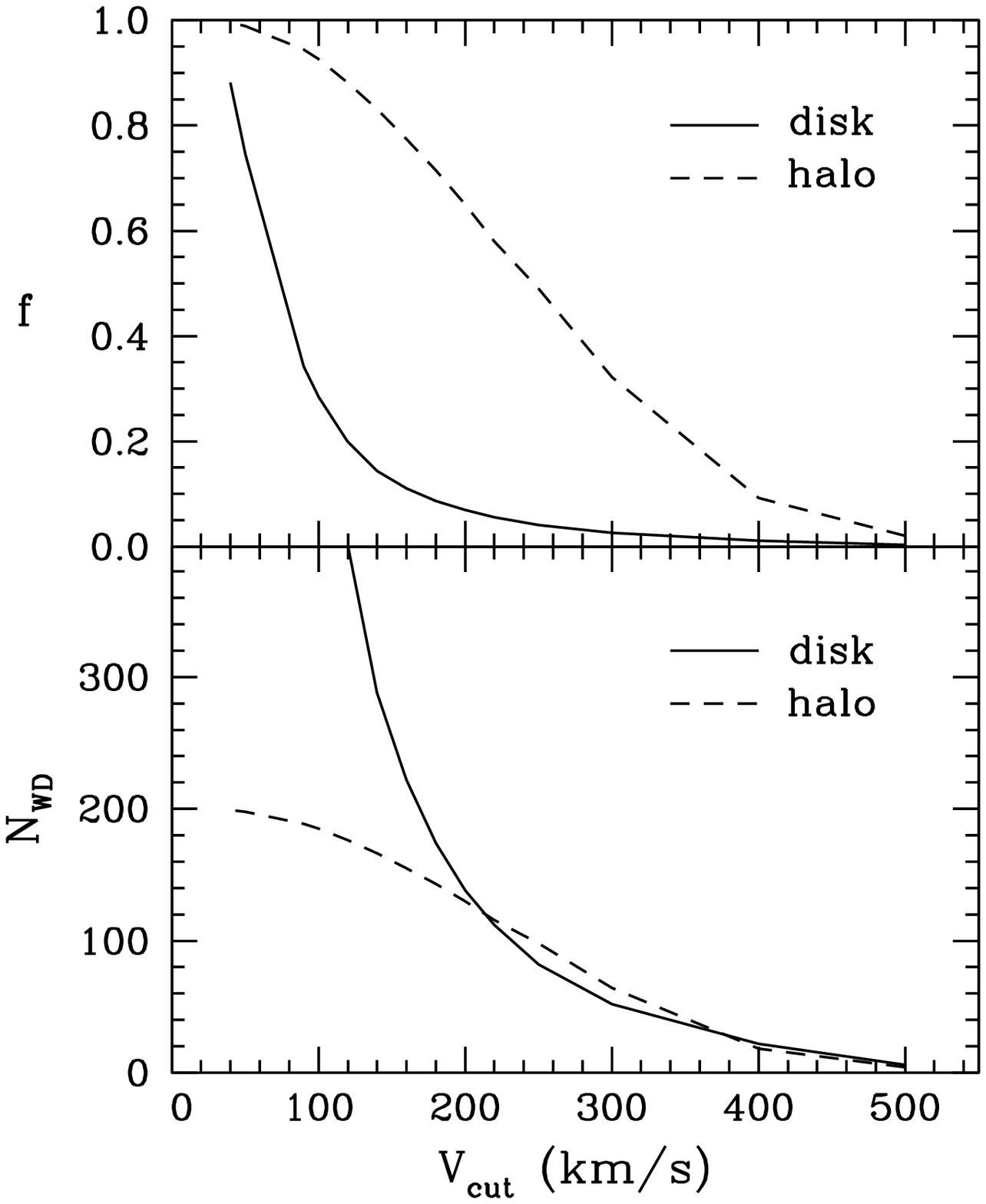}
\caption{Top panel:  fraction of white dwarfs discarded  as a function
         of  the  velocity  cut  for   both  the  disc  and  the  halo
         populations.  Bottom  panel:  total  number of  white  dwarfs
         discarded for both populations  as a function of the velocity
         cut.}
\end{figure}

Now, the  question is can  we reduce this  bias? The most  na\"ive and
straightforward method to do this is  to apply a velocity cut in order
to  separate  the  different  kinematical populations.   To  test  how
effective is  this widely  spread technique for  the case of  the disc
white  dwarf luminosity  function we  adopt the  most extreme  case in
which a 10\% contamination of the disc population by halo white dwarfs
is assumed.  The top panel of Fig.  9 shows the fraction of disc white
dwarfs (solid line) and halo  stars (dashed line) discarded using this
procedure as a function of the velocity cut, $V_{\rm cut}$. Obviously,
for  large velocity  cuts  the  fraction of  disc  stars discarded  is
totally negligible,  and increases as  the velocity cut  decreases, as
expected. The slope of the distribution turns out to be very steep for
velocity cuts  smaller than $\sim 120\;  {\rm km\;s^{-1}}$.  Regarding
the halo contamination it turns out that a velocity cut of $\sim 100\;
{\rm  km\;s^{-1}}$ discards $\sim  85$\% of  the contamination  of the
disc  population by  halo  white dwarfs.   The  resulting white  dwarf
luminosity function is almost  identical to that of the uncontaminated
population.  However,  this velocity cut can be  somewhat relaxed. For
instance,  adopting a velocity  cut of  $\sim 250\;  {\rm km\;s^{-1}}$
results in  a disc white dwarf luminosity  function totally equivalent
to  that  already shown  in  Fig.   8 for  the  case  in  which a  1\%
contamination  was   adopted.  We  recall   that  in  this   case  the
Choloniewski    method   gives    a   good    result,    whereas   the
$1/\mathcal{V}_{\rm max}$ estimator still  gives a biased one. In this
case the velocity cut must be reduced to $\sim 150\; {\rm km\;s^{-1}}$
in order to produce  acceptable results.  Thus, the necessary velocity
cut to remove  the halo contamination strongly depends  on the adopted
estimator,  being  considerably  larger  for  the case  in  which  the
Choloniewski method is employed.

In principle,  there is a wealth  of information in  the population of
white dwarfs  with large tangential  velocities, that is,  those which
were discarded from the complete sample using the previously described
procedure. Consequently, one  may think that it should  be feasible to
build   the  halo   white  dwarf   luminosity  function   using  these
stars. However, this is not  the case.  We illustrate the situation in
the lower panel of Fig.  9. In  this panel we show the total number of
white dwarfs that are discarded using different velocity cuts for both
the halo and the disc  populations.  These stars should be the natural
candidates to enter into a  pure halo sample.  However, even for large
velocity cuts the total number of  disc and halo white dwarfs are very
similar, thus preventing the derivation of a reliable halo white dwarf
luminosity function.   There are, however, more  sophisticated ways of
retrieving information from the  high velocity tail. These methods are
based in  artificial intelligence techniques (Torres et  al. 1998) and
require  the  incorporation  of  more  information  about  the  target
population like  colours and magnitudes  (among other characteristics)
of the stars of the sample.

\section{Conclusions}  

In this paper we have studied the biases introduced by the measurement
errors on the disc white  dwarf luminosity function. We have done this
for the  case in which a  small parallax--limited sample  of about 300
white dwarfs is used and for  a more interesting case in which a large
sample  of  white dwarfs  with  photometrically  derived distances  is
used. The first of these cases is representative of the current sample
from  which the  disc  white dwarf  luminosity  function is  obtained,
whereas the second case corresponds to the most recent sample of about
6000 white dwarfs obtained from the  SDSS Data Release 3. We have also
studied which luminosity function  estimator is more robust, analyzing
the behaviour of the two luminosity function estimators that have been
shown to perform best for the  case of the disc white dwarf luminosity
function when  no measurement errors  are taken into  account, namely,
the $1/\mathcal{V}_{\rm max}$ method and the Choloniewski method.  For
the case  of a small parallax--limited  sample we have  found that the
Lutz--Kelker bias is present and that it strongly affects the position
of  the cut--off  of  the  white dwarf  luminosity  function when  the
$1/\mathcal{V}_{\rm max}$ estimator is used. This is not the case when
the Choloniewski  method is used. In  this case, although  the bias is
present,  the  position of  the  cut--off  remains almost  unaffected.
However, we have also shown that using the appropriate techniques this
bias can  be removed and the  correct position of the  cut--off of the
disc white dwarf luminosity function can be retrieved, although at the
price of considerably increasing  the observational error bars. When a
large sample  of white dwarfs with photometric  parallaxes is studied,
the same behaviour is  found.  The $1/\mathcal{V}_{\rm max}$ method is
found to  be strongly biased,  providing an erroneous location  of the
cut--off  of  the white  dwarf  luminosity  function.   This bias  has
important consequences since a precise determination of the age of the
Solar neighbourhood requires an accurate location of the drop--off. We
have shown,  however, that  --- as it  was the  case in which  a small
parallax--limited sample  was used  --- the Choloniewski  method turns
out to be  rather insensitive to the measurement  errors and retrieves
with rather good accuracy the  position of cut--off of the white dwarf
luminosity function.

Finally, we  have also  studied the response  of both estimators  to a
potential  contamination  of  the  sample  of  the  disc  white  dwarf
population with halo  white dwarfs. We have found  that the effects of
such  contamination are quite  evident even  in the  case of  a modest
degree   of   contamination  (of   the   order   of   1\%)  when   the
$1/\mathcal{V}_{\rm max}$  method is  used.  The most  apparent effect
consists in a much shallower drop--off  in the case of a 1\% degree of
contamination. If  the contamination is higher  the drop--off actually
disappears. When the  Choloniewski method is used the  position of the
drop--off  is found  to be  much less  affected when  a 1\%  degree of
contamination is adopted,  although a new luminosity bin  at the faint
end of the white dwarf luminosity function --- exclusively due to halo
white  dwarfs ---  shows  up. This  bias  can be  removed  by using  a
velocity cut to cull from the sample only white dwarfs with relatively
small  tangential velocities. The  precise value  of the  velocity cut
depends on the  employed estimator. We have found  that a velocity cut
of $\sim 250\; {\rm km\;s^{-1}}$ works  fine for the case in which the
Choloniewski  method   is  employed.   For  the  case   in  which  the
$1/\mathcal{V}_{\rm max}$  is used this  velocity cut turns out  to be
considerably   smaller,   of   the   order   of   $\sim   150\;   {\rm
km\;s^{-1}}$. Consequently  more white  dwarfs are withdrawn  from the
original sample and the statistical significance of the resulting disc
white  dwarf luminosity  function  turns  out to  be  smaller in  this
case. All in  all, we find that the  Choloniewski method (Choloniewski
1986) is much  more robust  than the $1/\mathcal{V}_{\rm  max}$ method
against measurement  errors and a possible contamination  of the input
sample  by halo  white  dwarfs. Its  practical  implementation is  not
difficult and, moreover,  we have shown that it  retrieves an unbiased
estimate of  the position  of the cut--off  of white  dwarf luminosity
function, which by itself is a important reward.

\vspace{0.5 cm}

\noindent {\sl Acknowledgements.}  Part  of this work was supported by
the  MEC grants AYA05--08013--C03--01  and 02,  by the  European Union
FEDER funds and by the AGAUR.

\end{document}